# DYNAMIC SESSION KEY EXCHANGE METHOD USING TWO S-BOXES

Sohail Abid[1] and Shahid Abid[2]

[1]Department of Computing and Technology
IQRA University, Islamabad, Pakistan.
`rsohailabid@yahoo.com`
[2]Foundation University Institute of Engineering and
Management Sciences, Pakistan
`shahidkimail@yahoo.com`

*ABSTRACT*

*This paper presents modifications of the Diffie-Hellman (DH) key exchange method. The presented modifications provide better security than other key exchange methods. We are going to present a dynamic security that simultaneously realizes all the three functions with a high efficiency and then give a security analysis.*

*It also presents secure and dynamic key exchange method. Signature, encryption and key exchange are some of the most important and foundational Crypto-graphical tools. In most cases, they are all needed to provide different secure functions. On the other hand, there are also some proposals on the efficient combination of key exchange. In this paper, we present a dynamic, reliable and secure method for the exchange of session key. Moreover, the proposed modification method could achieve better performance efficiency.*

*KEYWORDS*

*S-Box key exchange, DSKE method, and three layer Security, Modified Diffie- Hellman key exchange.*

## 1. INTRODUCTION

This document describes, Network security issues are the highest priority of all network clients or users who want to secure their information and data. The network security issues are key concern for all businesses that would like to keep the verification, they are usually dealing with under restrict privacy. These security issues are not recent. When we want to send information from senders to recipients, the data transition has been prone to, the attacks in order to undercut the protection of useful data and information. In real meaning, these attacks are the outcome of strong opposition. As a matter of fact the network security problems are associated with computerized solutions. These computerized solutions are remarkable background. As long as information has some value, it is undeniably prone to attack.

Today the networks are rapidly expanding and the core issue is security. There are different levels of security like OS level, network level and session etc. Every one wants to secure his data, information and session. In this paper we are going to discuss few secure session key exchange methods and present a new method called DSKE. In a session key exchange each client need end to end session security and reliable communication.

Now a days each and every user is connected to the internet and internet is a link/ connection between users/clients, devices and servers. This link is insecure that is why security is a key issue. Today 90% communication is switched over data network and this change is rapidly growing.
The key exchange methods/ algorithms are one of the famous and known symmetric algorithms in the field of cryptography. There are different Session Key Exchange methods / algorithms Like Diffie-Hellman, Secure Hill Cipher Modifications and Key Exchange Protocol, Integration of Signature Encryption and Key Exchange, Secure Key Exchange and Encryption Mechanism for Ad Hoc Networks, Password Key Exchange Protocol. But every one has some weakness like insecure, huge calculation, slow and complex. We are trying to overcome these problems.

The Dynamic Session Key Exchange (DSKE) Method is computationally attractive as using multiplication of a key matrix. Our method has several advantages such as masquerading letter frequencies using matrix. The key exchange method is one of the well-designed ways of establishing secure communication between couple of users by using a session key. The session key, which is exchanged between two users, guarantee the secure communication for later sessions. The first practical key exchange method is proposed by Diffie-Hellman [1]. Since the introduction of key exchange method by Diffie-Hellman, a variety of versions and enhancement in key exchange method have been developed. In the line of key exchange method based key exchange mechanism achieved attention due to its complexity, dynamic security and wide range of applicability. In This method we take two S-Boxes S1 and S2. S1 is secret and S2 is chosen / taken from standard S2 box. S2 Standard box is open for all.  S1 is very secret; only two users understand this box. Using of these two S-Boxes, we can exchange session key between two users.

| 1a | 2a | 3a | 4a | 5a | 6a |
| 2b | 3b | 4b | 5b | 6b | 7b |
| 3c | 4c | 5c | 6c | 7c | 8c |
| 4d | 5d | 6d | 7d | 8d | 9d |
| 5e | 6e | 7e | 8e | 9e | ae |
| 6f | 7f | 8f | 9f | af | bf |

Figure 1a: S2 Box

| 1a | 2a | 3a | 4a | 5a | 6a |
| 2b | 3b | 4b | 5b | 6b | 7b |
| 3c | 4c | 5c | 6c | 7c | 8c |
| 4d | 5d | 6d | 7d | 8d | 9d |
| 5e | 6e | 7e | 8e | 9e | ae |
| 6f | 7f | 8f | 9f | af | bf |

Figure 1b: S2 Box

| 1a | 2a | 3a | 4a | 5a | 6a |
| 2b | 3b | 4b | 5b | 6b | 7b |
| 3c | 4c | 5c | 6c | 7c | 8c |
| 4d | 5d | 6d | 7d | 8d | 9d |
| 5e | 6e | 7e | 8e | 9e | ae |
| 6f | 7f | 8f | 9f | af | bf |

Figure 1c: S2 Box

International Journal of Computer Science, Engineering and Applications (IJCSEA) Vol.1, No.6, December 2011

## 2. PROPOSED DESIGN

There are many session key exchange methods and algorithms but the most popular method use private and public key. In conventional Public Key Infrastructure (PKI), there is an essential to provide guarantee to the client about the relationship between a public key and the public key authority of the corresponding private key. In practice there are many challenges which are facing PKI like distribution of certificates, Storage and revocation. In order to solve the above problem, certificateless Public Key Cryptography (CL-PKC) was introduced.

The new prototype called Self-Generated-Certificate Public Key cryptography without pairing (SGC-PKC) proposed by J. Lai and W. Kou [2] to protect the above attack while preserving all advantages of Certificateless Public Key Cryptography. In [3], this paper writer proves that Lai and Kou's method cannot defend against a type of middle attack. In order to solve this problem to propose a new SGC-PKE method by giving small difference to the original method.

Therefore and propose a rescue SGC-PKE scheme by giving little change to the original scheme [4]. The USA Department of Defence developed HAIPE (The High Assurance Internet Protocol Encryptor) having compliant gateways to communicate securely over un-trusted networks. In [5] this paper created automated security association by using Internet Key Exchange (IKE) and HIPEs mutually.

In Off-line password guessing the hacker first guess password and then verifies it online. In this method the hackers bypass the server. Therefore server can not verify the attack. Ding and Hoster proposed [6], in their paper on line and offline guessing attacks on Stener's Protocol. There are many password based efficient key exchange protocols. The Bellovin and Merrit first proposed (PAKE) two Party key exchange protocols [7]. After that Steiner et al [8] in his paper proposed the 3-party protocol. Two type of improved three party protocol proposed by Lin et al [9]. One type used with server and other without server. Chang and Chang[10] proposed without server novel 3-party encrypted key exchange method and claim that this method is efficient and secure. But Yoon and Yoo claim an undetectable password guessing attack on their method [11] and proposed new method which avoid these attacks. Further Lo, Yes proposed an enhance method which handles undetectable password guessing attacks [12]. Some new techniques are also introduced like "Security Verification for Authentication and Key Exchange Protocols, Revisited"[17], "New Framework for Efficient Password-Based Authenticated Key Exchange"[16], "Multi-Factor Password-Authenticated Key Exchange"[19] and "An Efficient Four-Party Key Exchange Protocol for End-to-End Communications"[20].

We proposed and alternative method for key exchange. In our method both users take any 3x3 or 4x4 or 5x5 box from S2 Box. Then select 3x3 or 4x4 or 5x5 S1 box which is hidden. After the selection / chosen of S1 and S2 Boxes, both parties decide two large prime numbers P and Q and third number n which is small. All the three numbers are secret. Both parties create their S1 Box using this method.

| S1 | | | | | |
|------|------|------|------|------|------|
| S100 | S101 | S102 | S103 | S104 | S105 |
| S110 | S111 | S112 | S113 | S114 | S115 |
| S120 | S121 | S122 | S123 | S124 | S125 |
| S130 | S131 | S132 | S133 | S134 | S135 |
| S140 | S141 | S142 | S143 | S144 | S145 |
| S150 | S151 | S152 | S153 | S154 | S155 |

Figure 2: S1 Box

P=5, Q=29 and n=3

```
S100 = P mod Q          5   mod 29 = 5
S101 = Pn mod Q         5^3  mod 29 = 125 = 9
S102 = Pn+1 mod Q       5^4  mod 29 = 625= 16
S103 = Pn+2 mod Q       5^5  mod 29 = 3125= 22
S104 = Pn+3 mod Q       5^6  mod 29 = 15625= 23
S105 = Pn+4 mod Q       5^7  mod 29 = 78125= 28

S110 = P – n mod Q      5 - 3 mod 29 = 2
S111 = P – 2n mod Q     5-2(3)  mod 29 = -1 =28
S112 = P – 3n mod Q     5- 3(3)  mod 29 = -4 =25
S113 = P – 4n mod Q     5- 4(3)  mod 29 = -7 =22
S114 = P – 5n mod Q     5- 5(3)  mod 29 = -10 =19
S115 = P – 6n mod Q     5- 6(3)  mod 29 = -13 =16

S120 = P + n mod Q      5+3 mod 29 =8
S121 = P + 2n mod Q     5+2(3) mod 29 = 11
S122 = P + 3n mod Q     5+3(3) mod 29 = 14
S123 = P + 4n mod Q     5+4(3) mod 29 = 17
S124 = P + 5n mod Q     5+5(3) mod 29 = 20
S125 = P + 6n mod Q     5+6(3) mod 29 = 23

S130 = P x 2 - n mod Q  5x2 - 3 mod 29 = 7
S131 = P x 3 - n mod Q  5x3 - 3 mod 29 = 12
S132 = P x 4 - n mod Q  5x4 - 3 mod 29 = 17
S133 = P x 5 - n mod Q  5x5 - 3 mod 29 = 21
S134 = P x 6 - n mod Q  5x6 - 3 mod 29 = 27
S135 = P x 7 - n mod Q  5x7 - 3 mod 29 = 32=3

S140 = P + 2 - n mod Q  5+2 - 3 mod 29 = 4
S141 = P + 3 - n mod Q  5+3 - 3 mod 29 = 5
S142 = P + 4 - n mod Q  5+4 - 3 mod 29 = 6
S143 = P + 5 - n mod Q  5+5 - 3 mod 29 = 7
S144 = P + 6 - n mod Q  5+6 - 3 mod 29 = 8
S145 = P + 7 - n mod Q  5+7 - 3 mod 29 = 9

S150 = P x 2 + n mod Q  5x2 + 3 mod 29 = 13
S151 = P x 3 + n mod Q  5x3 + 3 mod 29 = 18
S152 = P x 4 + n mod Q  5x4 + 3 mod 29 = 23
S153 = P x 5 + n mod Q  5x5 + 3 mod 29 = 28
S154 = P x 6 + n mod Q  5x6 + 3 mod 29 = 33= 4
S155 = P x 7 + n mod Q  5x7 + 3 mod 29 = 38= 9
```

S1

| 5  | 9  | 16 | 22 | 23 | 28 |
|----|----|----|----|----|----|
| 2  | 28 | 25 | 22 | 19 | 16 |
| 8  | 11 | 14 | 17 | 20 | 23 |
| 7  | 12 | 17 | 21 | 27 | 32 |
| 4  | 5  | 6  | 7  | 8  | 9  |
| 13 | 18 | 23 | 28 | 4  | 9  |

Figure 2: S1 Box

International Journal of Computer Science, Engineering and Applications (IJCSEA) Vol.1, No.6, December 2011

If we check S1 Box there are may repeated numbers. If we take large prime numbers P and Q then there is very less repeated numbers. Choose S1 Box has no repeated number.
Now we take 3x3 S1 and S2 Boxes.

| S1 | | | | S2 | | |
|---|---|---|---|---|---|---|
| 17 | 20 | 23 | | 3b | 4b | 5b |
| 21 | 27 | 32 | | 4c | 5c | 6c |
| 7  | 8  | 9  | | 5d | 6d | 7d |

Both parties understand S1 and S2 boxes. When A send 17 to B, A send 3b to B. B receives 3b and understand as 17.

## 3. SECURITY ANALYSIS

The correctness of this method can be easily seen from the description of the method, it do synthetically achieve the goals of DH key agreement. And the security of the encryption depends on the p, q and n are at layer 1. The selection of S1 Box is at layer2. The selection S2 Box is at layer3. The p, q and n are selected for long time unless both parties feel insecurity of p, q and n. But layer 2 and layer 3 changes every session. In this method the security is presented in three layers.

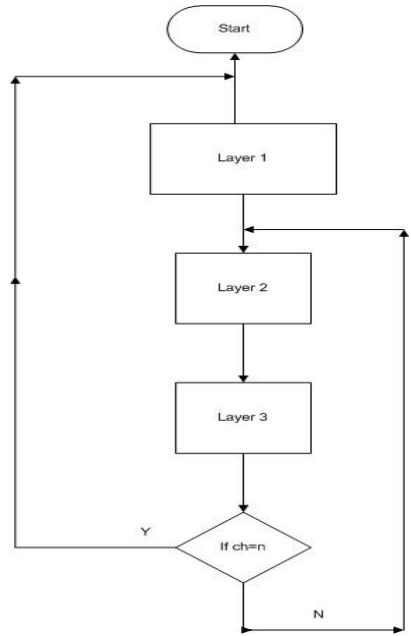

Figure 3: Security Layers

### 3.1. Case I

If layer 2 and layer 3 are broken or hacked in a session. The rest of the part is open. But the next session remains secure. Because layer 2 and layer 3 will be change in the next session.

International Journal of Computer Science, Engineering and Applications (IJCSEA) Vol.1, No.6, December 2011

Let us see this scenario Bob wants to send 21 to Alice. On the other side Alice receives 4c and after using of S boxes we will get 21. If this session hacked the next session will be secured because they chose another number.
In First Session

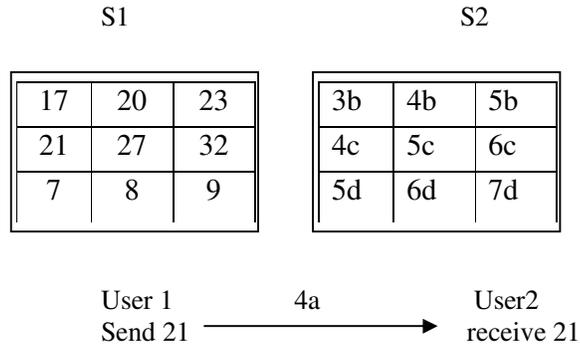

In Second Session the S1 box changes.

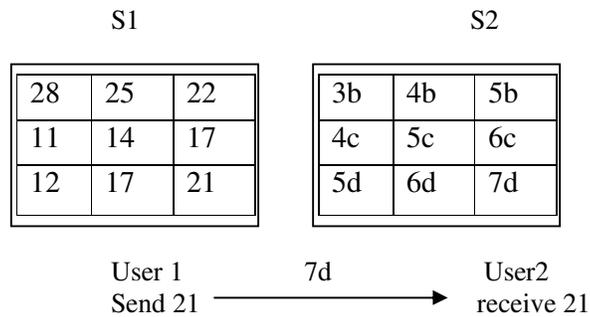

### 3.2. Case II

In case the layer 1 and layer 2 are broken or hacked in a session. The session is secure. In next session layer2 and layer 3 changes so the next session will be secure.
In First Session

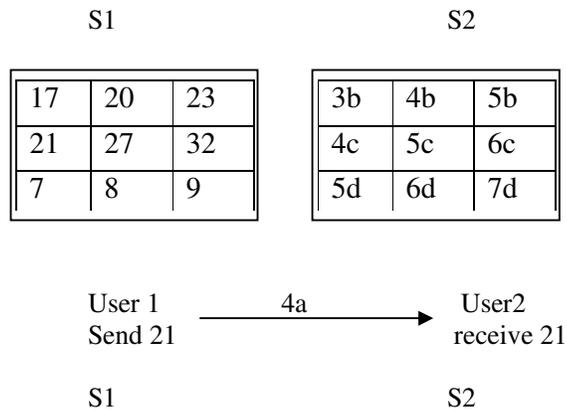
S1                                    S2
100

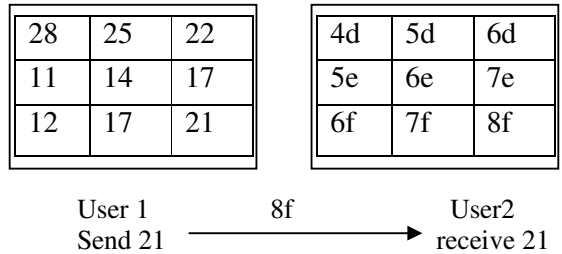

## 3.3. Case III

In case only one layer is broken .The session is secure.
In the light of above three cases we analyze that if we decrease the time of changing layer1. The security of session key is highly secure. In this paper we present such security that depends upon both users.

## 4. APPLICATION AREAS

This method is used in any computer and network security area. Specially design for session key. This session key exchange method is dynamic, secure and fast. The utilization of this method is in every security related paradigm some of them are as under:

### 4.1. Clouds Security

Cloud computing is a service like shared databases, software, resources and information provided to devices and computers over an internet. Cloud Computing has many types some of them are public clouds 2, private clouds 3. In other words cloud computing is system in which clients can access or user remotely shared resources, databases, software and information. These services are very effective for businessmen, and layman. But there is a problem of security during a session [13, 14,15].

### 4.2. Grid Security

The term grid computing is used for collectively shared resources of PC from multiple domains and to achieve a common objective. GRID (Global Resource & Information Directory) is planned to make the Internet a secure and better place for a single, accurate and state-of-the-art for all users. There are three broad types of threats for smart grid Computing: to take down the grid, to compromise data confidentiality and to steal electrical service on that PC. End-to-end advanced security is required for grid computing [15].

### 4.3. Distributed Data Security

In distributed computing more than one computer share their resources like processing power, hard disk (storage) and RAM to achieve their common goal. Distributed computing is to solve large computational problems in such a way that a group of network computers achieve their common goal for their work. Security of whole communication is the main issue, which is weak part.

### 4.4. Wireless Network Security

WiFi is a local WLAN (Wireless LAN), the coverage of WLAN is limited near about 40 meters. In WLAN all computers or devices are connected to each other over a wireless / radio

International Journal of Computer Science, Engineering and Applications (IJCSEA) Vol.1, No.6, December 2011

connectivity. WiMax is a WWAN (Wireless WAN) is connectivity between two cities over a radio link or satellite link. In wireless environment the major importance is mobility and major threat is security.

### 4.5. Next Generation Network Security

The next generation network is actually a wireless network. In which the main concept is mobility and service of network and all other services are available every where. Network security issues are the top priority of all who want to protect their data. Usually, network security issues are of major concern for all businesses that want to keep the affirmation they are usually dealing with under strict confidentiality. Session Initiation protocol SIP plays important role in NGN [18]. As a matter of fact the network security problems are associated with computerized solutions. These computerized solutions are remarkable background. There is lot of researcher work on security and security is a core issue of wireless networks, distributed computing, Cloud Computing, Grid and NGN.

## 5. CONCLUSIONS

In this paper we present numerous Diffie-Hellman (DH) key exchange method. The DH modifications providing enhanced security and encryption quality than known ones. A method for secure key exchange similar to DH, but using dynamic session key exchange method instead of less secure and time-consuming calculation is also proposed. Our proposed method is more reliable then other key exchange methods.

## 6. ACKNOWLEDGEMENT

We are greatly acknowledge the kind supervision of Dr. Ismail Shah, who taught me the subject Advanced Network Security and encouraged me for this work and to write this paper.

## REFERENCES

[1]  W. Diffie and M. Hellman, "New Directions in cryptography", IEEE Transactions on Information theory, Vol 22 ,no. 6 , pp 644-54, (1976).

[2]  Junzuo Lai, Weidong Kou. Self-Generated-Certificate Public Key Encryption Without Pairing[C]. PKC 2007, Beijing, China, April 16-20, 2007. Springer-Verlag, 2007, LNCS 4450, pp. 476-489.

[3]  Xu An Wang, Xiaoyuan Yang and Yiliang Han. Cryptanalysis of Self-Generated-Certificate Public Key Encryption without Pairing in PKC07[EB/OL], Cryptology ePrint Archive: Report 2008/191, 2008. http://eprint.iacr.org/2008/191.

[4]  Hua Jiang, Rui Zhang and Yongxing Jia, "Authenticated Key-Exchange Scheme Based on SGC-PKE for P2PSIP", NSWCTC '10 Proceedings of the 2010 Second International Conference on Networks Security, Wireless Communications and Trusted Computing – IEEE Computer Society, Vol 02, pp 352-    356, (2010).

[5]  ZHANG Chuan-fu#1, YU Jiang#2, SunWan-zhong#3, SU Jin-hai#4, "Internet Key Exchange Protocol Simulation of HAIPE in Security Network", 2010 International Conference on Cyber-Enabled Distributed Computing and Knowledge Discovery.

[6]  Y. Ding and P. Hoster, "Undetectable Online password guessing attacks", ACM operatinf system review, vol 29, no 4,pp 77-86 (1995)

[7]  SM. Bellovin and M. Merrit, "Encrypted key exchange: password based protocols secure against dictionary attacks". IEEE sysmposium on re-search in security and privacy, IEEE Computer society press :72-84,(1992).

[8]  M. Steiner and G. Tsudik, M. Waidner "Refinement and extention of encrypted key exchange", ACM Operating Systems Review, vol 29, no 3, pp 22-30, ( 1995).

[9]  CL. Lin, HM. Sun, M. Steiner, T. Hwang " Three-party excrypted key exchange without server public Keys" IEEE Communication letters, vol 5, no.12,pp 497- 9 , (2001).

[10] CC. Chang and YF. Chang, "A novel three party encrypted key exchange protocol", Computer Standards and Interfaces, vol 26 , no 5, (pp 471-6),(2004).

[11] EJ. Yoon and KY. Yoo, "Improving the novel three-party encrypted key exchange protocol", Computer Standards and Interfaces, 30:309-314 , (2008).

[12] N.W.Lo,Kuo-Hui Yeh, "Cryptanalysis of two three-party Encrypted key exchange protocols", In press, computer standards and interfaces.

[13] Jean Bacon, David Evans etl, "Enforcing End-to-End Application Security in the Cloud", Indranil Gupta and C. Mascolo (Eds.): Middleware 2010, LNCS 6452, pp. 293–312, 2010, © IFIP International Federation for Information Processing 2010.

[14] Benoît Bertholon_, Sébastien Varrette _ and Pascal Bouvry, "CERTICLOUD: a Novel TPM-based Approach to Ensure Cloud IaaS Security", 2011 IEEE 4th International Conference on Cloud Computing.

[15] Pankaj Goyal, "Application of a Distributed Security Method to End-2-End Services Security in Independent Heterogeneous Cloud Computing Environments", 2011 IEEE World Congress on Services.

[16] Adam Groce and Jonathan Katz, "New Framework for Efficient Password-Based Authenticated Key Exchange", CCS'10, October 4–8, 2010, Chicago, Illinois, USA. Copyright 2010 ACM 978-1-4503-0244-9/10/10.

[17] Haruki Ota, Shinsaku Kiyomoto and Toshiaki Tanaka, "Security Verification for Authentication and Key Exchange Protocols, Revisited", 2010 IEEE 24th International Conference on Advanced Information Networking and Applications Workshops.

[18] Vijay K. Gurbani, Vladimir Kolesnikov," Work in progress: A secure and lightweight scheme for media keying in the Session Initiation Protocol (SIP)", IPTComm 2010, 2-3 August, 2010 Munich, Germany Copyright 2010 ACM.

[19] Douglas Stebila, Poornaprajna Udupi and Sheueling Chang, "Multi-Factor Password-Authenticated Key Exchange", Copyright c 2010, Australian Computer Society, Inc. This paper appeared at the Australasian Information Security Conference (AISC), Brisbane, Australia. Conferences in Research and Practice in Information Technology (CRPIT), Vol. 105, Colin Boyd and Willy Susilo, Ed.

[20] Wei-Kuo, Chiang and Jian-Hao Chen,"TW-KEAP: An Efficient Four-Party Key Exchange Pro tocol for End-to-End Communications", SIN'11, November 14-19, 2011, Sydney, Australia. Copyright 2011 ACM 978-1-4503-1020-8/11/11

**Authors**

Sohail Abid: (Mobile No: +92-321-5248497)
Sohail Abid Student of MS (TN) at IQRA University Islamabad and working as System Administrator at Foundation University Institute of Engineering and Management Sciences.

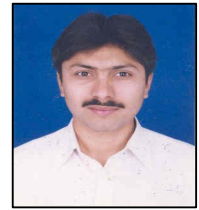

Shahid Abid (Mobile No: +92-333-5656413)
Shahid Abid having Master in Computer Science and working as Assistant System Administrator at Foundation University Institute of Engineering and Management Sciences.

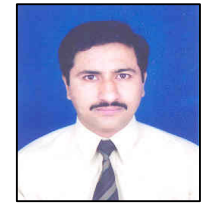